\documentstyle[times,pramana,epsf,floats]{ias}
\begin{document}
\mark{{Synchronization and Information...}{Kakmeni and Baptista}}
\title{Synchronization and Information Transmission in Spatio-Temporal Networks
  of Deformable Units} \author{F. M. Moukam Kakmeni$^{1,2}$ and M. S.
  Baptista$^{1,3}$} \address{$^1$Max-Planck-Institut f\"ur Physik
  komplexer Systeme, N\"othnitzerstr. 38, D-01187 Dresden,
  Deutschland} \address{$^2$Department of Physics, Faculty of
  Science, University of Buea, P. O. Box 63 Buea, Cameroon}
\address{Centro de Matem\'atica da Universidade do Porto, Rua do Campo
Alegre 687, 4169-007 Porto, Portugal}

\abstract{We study the relationship between synchronization and the
  rate with which information is exchanged between nodes in a
  spatio-temporal network that describes the dynamics of classical
  particles under a substrate Remoissenet-Peyrard potential. We also
  show how phase and complete synchronization can be detected in this
  network. The difficulty in detecting phase synchronization in such a
  network appears due to the highly non-coherent character of the
  particle dynamics which unables a proper definition of the phase
  dynamics. The difficulty in detecting complete synchronization appears
  due to the spatio character of the potential which results in an
  asymptotic state highly dependent on the initial state.}

\maketitle
\section{Introduction}
The sine-Gordon potential and similar others have been used to
model the dynamics of many systems in physics, biology and
engineering\cite{Braun1,Remoissenet1,Remoissenet2,Djuidje1,Kofane1,Nguenan1,Nana1,Yamgoue1}.
However, in real physical systems, the shape of the substrate
potential can deviate from the standard one with a direct
incidence on the stability properties of the system. In physical
situations, such as charge-density waves, Josephson junctions, or
crystals with dislocations, the application of the standard
sine-Gordon model becomes too restrictive.  In recent years, a
number of potentials whose shapes can be turned at wish have
appeared in the literature of nonlinear dynamical systems
\cite{Remoissenet1,Remoissenet2,Djuidje1,Kofane1,Nguenan1,Nana1,Yamgoue1}.
These more realistic potentials certainly provide richer insights
onto the physics of reals systems than what is predicted using the
conventional, rigid models such as the sine-Gordon,
double-sine-Gordon and $\phi^4$ potentials. In particular, we can
expect a more rich and complex synchronization phenomena in models
of nonlinear oscillators involving them.

The purpose of the present paper is to study networks formed by
oscillators under realistic shape deformable potentials. To model the
network, we use the Remoisnet-Payrar potential, which has been
extensively used in the literature to describe the disturbance of the
sinusoidal shape of the substrate periodic potential of the Sine
Gordon equation\cite{Braun1,Remoissenet1,Remoissenet2}.

We are mainly interested in the complex relationship between
synchronization and transmission of information.  By
synchronization, we mean complete synchronization (CS)
\cite{Boccaletti1,pecora,yamapi,Boccaletti2} and chaotic phase
synchronization (PS) \cite{Kurths1}. The information point-of-view
will be provided by the procedure described by Baptista {\it et
al.} in Refs. \cite{murilo}. As we shall show synchronization and
information are directly related in such an active network. The
larger the synchronization is the larger the rate with which
information is exchanged between nodes in the network, the so
called mutual information rate (MIR).

Such relationship can be experimentally explored when one needs to
observe how  nodes are attached to each other in a real network.
For situations where the nodes of the network are neither
completely synchronous nor phase synchronous, the MIR provides one
the level of connectivity. In addition, the MIR limits the amount
of information that can be retrieved in some point of the network
about an arbitrary external stimulus.

Due to the spatio character of the studied network, both
approaches, the ones  in Refs.
\cite{Boccaletti1,pecora,yamapi,Boccaletti2} and the ones in Refs.
\cite{murilo} might face difficulties to be implemented and this
work resolves many of them. In particular, we study networks which
have node trajectories departing from randomly initial conditions.
That creates a situation similar to the one observed in networks
constructed with nodes presenting different parameters, when the
methods in Ref. \cite{pecora} should be used with precaution.

Note that a quite number of physical objects allowing a model description with
the aid of the sine-Gordon equation are known: arrays of forced damped
pendula, vortices in long Josephson junctions, charge-density waves in
quasi-one-dimensional conductors etc...\cite{Braun1,Remoissenet1}. For real
physical systems, the account of various disturbances and of a more complex
character of atomic interactions breaks the exact integrability of the initial
sine-Gordon equation, leaving the possibility for describing the system
dynamics in terms of the same quasi-particles which now interact with each
other.

The rest of the paper is organized as follows: in Sec.
\ref{secaoII} we explore the dynamics of the network in
consideration and analyze the effect of the deformability
parameter in the substrate potential on the stability
synchronization of the network. In Sec. \ref{secaoIII} we analyze
phase synchronization in such networks, and Sec. \ref{secaoIV} is
devoted to the study of information transmission within the nodes
of the network. Finally, we present the conclusions in Sec.
\ref{secaoV}.

\section{Synchronization dynamics of the networks}\label{secaoII}
\subsection{Description of the networks}\label{secaoIIa}

We first investigate the dynamical properties of a single particle in
a deformable substrate potential. If we define the variable $x$ as the
displacement of the particle in the potential well, then the equation
of motion describing its dynamics reads

\begin{equation}
\label{mouk01}
  \ddot{x}+\lambda \dot{x}+\omega^2
   \displaystyle\frac{\partial V(x,r)}{\partial x } =  \eta_0\cos\Psi
   t.
\end{equation}
\noindent In this work we consider the following fixed set of
parameters $\lambda=0.01, \omega=1, \eta_0=0.19$. The parameters
$\Psi$ and $r$ will be varied.

Recall that $x$ is the coordinate variable which characterizes the
behavior of the particle in the potential well $V(x,r)$. The new
issues of our model under consideration are the following: we
apply an AC force $\eta(t)=\eta_0\cos\Psi t$ to the particle and
assume also the external viscous damping with a coefficient
$\lambda$. In this work, $V(x, r)$ is a nonlinear potential with a
deformable shape introduced by Remoissenet-Payrard to study the
coherent structure in a network formed by a similar system. There
are many versions of this potential, but we concentrate our
analyses on the most general case defined as
\cite{Remoissenet1,Remoissenet2,Djuidje1,Kofane1,Nguenan1,Nana1,Yamgoue1}
\begin{equation}
\label{mouk02}
  V(x,r)=(1-r)^2\displaystyle\frac{1-\cos x}{1+r^2+2r\cos x}
\end{equation}
\begin{figure}[htbp]
\epsfxsize=8cm \centerline{\epsfbox{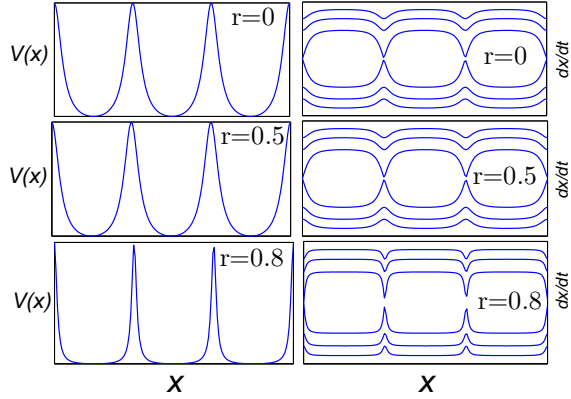}} \caption{Form
of the potential as a function of $r$ and the corresponding
periodic orbits for a free particle ($\eta_0$=0 and $\lambda=0$).
The pictures in the left (right) column shows $x$ vs. $V(x,r)$
($x$ vs. $\dot{x}$).} \label{Fig.1}
\end{figure}
\noindent
where the deformability parameter $r$ fulfills the condition $|r|<1$.

The advantageous feature of this potential can be summarized in the
fact that it reproduces the sine Gordon ($r=0$) while avoiding most of
it shortcomings. A shape of broad wells separated by narrow barriers
can be obtained for $r>0$ and for $r<0$, a shape of deep narrow wells
separated by broad gently sloping barriers can be obtained.

Figure \ref{Fig.1} shows the form of the potential and the
corresponding phase plane as a function of the parameter $r$, for
$r>0$. One can observe that the larger the parameter $r$ is, the
flatter the bottom of the potential.

In real physical systems, such potential can be produced by the
interaction of an adatom with substrate atoms, where the parameter
$r$ could account for the temperature or pressure dependence, or
for the geometry of the surface of the metallic surface. It can be
calculated from the first principles as described in Refs.
\cite{Braun1,Remoissenet1,Djuidje1} (and reference therein).
However it is more reliable to determine the parameter $r$ from
experimental data. Estimates for e.g., a H/W adsystem (hydrogen
atoms absorber on a tungsten surface), yield
$r\approx-0.3$\cite{Braun1,Remoissenet1,Djuidje1}
\begin{figure}[htbp]
\epsfxsize=8cm \centerline{\epsfbox{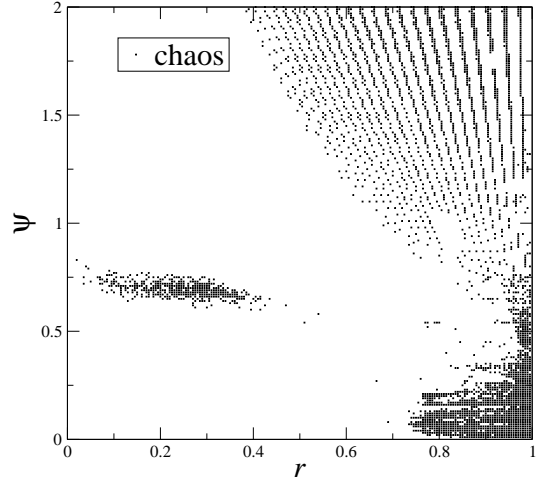}}
\caption{Parameter space plot of the frequency $\psi$ and the
  deformability parameter $r$. Points represent chaotic behavior
  (positive KS-entropy and continuous Fourier spectrum).}
\label{Fig.2}
\end{figure}

Typically, if periodic oscillators are subjected to a periodic force,
different phase-locking phenomena as well as chaos may be observed.
And chaotic oscillators when subject to a periodic force give rise to
a series of bifurcation phenomena.

Figure \ref{Fig.2} shows the parameter space diagram of the
oscillator in Eq.  (\ref{mouk01}). Points (blank space) indicate
values of the frequency $\Psi$ and the deformability parameter $r$
for which the oscillator in Eq.  (\ref{mouk01}) is chaotic
(periodic).

The $(r, \Psi)$ space is characterized by the predominance of
periodic solutions. The chaotic solutions appear only for the
value of the deformability parameter approaching the limit $1$.
However for high frequency, the chaotic motion appears earlier,
that is at $r \simeq 0.4$. For larger $r$ and $\Psi$ the parameter
space presents a complex pattern whose chaotic regions appear
side-by-side with periodic regions.  For the specific narrow band
of the frequency $\Psi$ around $0.70$ and $0.75$ a deep band of
chaotic motion can be found for $r$ between $0.1$ and $0.4$.  This
confirms the chaotic behavior of deformable models systems as
first suggested in references \cite{Djuidje1,Nana1,Yamgoue1}.

We now consider a network of $N$ dynamical units of oscillators
described by equations (\ref{mouk01}) and (\ref{mouk02}). The
governing equation for the network is given by:
\begin{eqnarray}\label{mouk03}
\dot{p_i}&=& n_i \nonumber\\
\dot{n_i}&=&-\lambda n_i -\omega^2
   \displaystyle\frac{\partial V(p_i,r)}{\partial p_i }+\eta_0\cos\Psi
   t\\
   &&+g_l(p_{i+1}-2p_{i}+p_{i-1}) \:\:\:\:\:\:\: \textrm{with} \:\:\:\:\:\: i=1,2...,N\nonumber
\end{eqnarray}
where $V(p_i, r)$ is given by Eq. (\ref{mouk02}). The constant
parameter $g_l$ determines the strength of the coupling and $N$
the number of oscillators coupled. This equation is known as the
Frenkel-Kontorova(FK) model with harmonic interaction and
non-sinusoidal substrate potential. It has been extensively
studied in the research of static characteristics of kinks
(topological solitons) such as the effective mass, shape, and
amplitude of the Peierls potential, the interaction energy of
kinks, and the creation energy of kink-antikink pairs. The
applicability of the extended Frenkel-Kontorova model for
describing diffusion characteristics of a quasi-one-dimensional
layer adsorbed on a crystal surface has also been discussed in
Ref. \cite{Braun1}. For real physical systems, the account of
various disturbances and of a more complex character of atomic
interactions break the exact integrability of the initial Sine
Gordon equation, leaving the possibility for describing the system
dynamics in terms of the same quasi-particles which interact with
each other. This interaction, which is due to the departure from
complete integrability, results in the following effects. The
kolmogorov-Sinai entropy becomes nonzero and the Fourier spectrum
of excited states of the system becomes continuous. Both
characteristics of chaos.

\subsection{Stability of the synchronization}\label{secaoIIb}

\noindent Our analysis will be limited to networks of identical
units. Since the $N$ systems are identical, it exists an exactly
synchronized solution of Eq. (\ref{mouk03}), and the
synchronization manifold is defined by $\mathcal{M}$=$\{p_1=p_2=
.... =p_N=p_s; n_1=n_2=....=n_N=n_s\}$.

In the study of synchronization, a very relevant problem is to
assess the conditions for the stability of the synchronous
behavior for the networks and for the coupling configuration. The
master stability function approach was originally introduced for
arrays of coupled oscillators \cite{pecora}, and it has been
latter extended to the case of complex networks of dynamical
systems \cite{Boccaletti1,Boccaletti2}. To use this, let us
consider $N$ coupled dynamical units, each of them giving rise to
the evolution of 2-dimensional vector fields $x_i$ ruled by a
local set of ordinary differential equations $\dot{x}_i =
\textbf{F}(x_i)$. The equations of motion using the new variable
can be written as
\begin{eqnarray}
\label{mouk04} \dot{x}_i=\textbf{F}(x_i)+g_l\sum_{j=1}^N
G_{ij}\textbf{H}(x_j), \qquad i=1,2,...,N,
\end{eqnarray}
where $\dot{x}_i=\textbf{F}(x_i)$ governs the local dynamics of
the $i$th node. $x_i=[p_i,n_i]^T,$ and $\textbf{F}(x_i)=\left[n_i,
-\lambda n_i -\omega^2
   \frac{\partial V(p_i,r)}{\partial p_i }+\eta_0\cos\Psi
   t \right]^T$ with $V(p_i, r)$ as in Eq.(2),
the output function $\textbf{H}(x_i)$ is a vectorial function
defined trough the matrix
$\mathbf{ E }$=$\left (
\begin{array}{cc}
0 &0\\
1 &0
\end{array}
\right )$
by $\textbf{H}(x_i)=$ $\mathbf{ E }$ $x_i$ , and $G(t)$ is a symmetric
Laplacian matrix ($\sum_jG_{ij}=0$) describing the networks connection
and given by
\begin{displaymath}
\mathbf{ G }=\left (
\begin{array}{ccccc}
-2&1& 0& \ldots &1\\
1 &-2&1& \ldots &0\\
0&1&-2&\ldots&0\\
\vdots &\vdots&\vdots&\ddots&1\\
1&0&\ldots &1&-2
\end{array}
\right )
\end{displaymath}

The stability of the synchronization state can be determined from
the variational equations obtained by considering an infinitesimal
perturbation $\delta x_i$ from the synchronous states, $p_i=\delta
p_i+p_s$, $n_i=\delta n_i+n_s$. The equations of motion for the
perturbation $\delta x_i$ can be straightforwardly obtained by
expanding the Eq. (\ref{mouk04}) in Taylor series of first order
around the synchronized state which gives
\begin{eqnarray}
\label{mouk05} \delta {\dot x}_i&=& D \textbf{F}(x_s)\delta
x_i+g_l\sum_{j=1}^N G_{ij}D \textbf{H}(x_s)\delta x_i,
 \:\:\:\:\:\:\:\: i=1,2,...,N,\nonumber\\
&=&\sum_{j=1}^N \left[ D \textbf{F}(x_s)\delta_{ij}+g_l G_{ij}D
\textbf{H} (x_s)\right]\cdot\delta x_i, \:\:\:\:\:\: i=1,2,...,N,
\end{eqnarray}
where $D\textbf{F}$ and $D\textbf{H}$ are the Jacobians of the
vector field and the output function respectively.

Equation (5) is referred to as the variational equation and is
often the starting point for stability determination. This
equation is rather complicated since given arbitrary coupling $G$
it can be quite high dimensional. However, we can simplify the
problem by noticing that the arbitrary state $\delta x_i$ can be
written as $\delta x_i=\sum_{
  i=1}^N \textbf{v}_i \bigotimes \xi_i(t)$ with $\xi_i(t)=(\xi_{
  1,i},\xi_{ 2,i})$ where $\gamma_i$ and $\textbf{v}_i$ are the set of
real eigenvalues and the associated orthogonal eigenvector of the
matrix $G$ respectively, such that $G\textbf{v}_i=\gamma_i
\textbf{v}_i$ and $\textbf{v}_i^T \textbf{v}_i = \delta_{ij}$. By
applying $\textbf{v}_i^T(t)$ (and $v_i$) to the left (right) side
of each term in Eq. (\ref{mouk05}) one finally obtains a set of N
blocks for the coefficients $\xi_i(t)$. The first term with the
Kronecker delta remains the same. This results in a variational
equation in the eigenmode form
\begin{eqnarray}
\label{mouk06} \dot \xi_k=\left[ D \textbf{F}(x_s)+g_l \gamma_k D
\textbf{H}(x_s)\right]\xi_k, k=0,1,2,...,N-1,
\end{eqnarray}

We recall that $\gamma_k$ are the eigenvalues of $G$, and are
given by $\gamma_k=-4\sin^2(\pi k/N)$ for the diffusive coupling
\cite{pecora}. Note that each equation in Eq. (\ref{mouk06})
corresponds to a set of 2 conditional Lyapunov exponents
$\lambda_k^j$ (j=1,2) along the eigenmode corresponding to the
specific eigenvalue $\gamma_k$. For $k=0$, we have the variational
equation for the synchronization manifold $(\gamma_0=0)$ and its
maximum conditional Lyapunov exponent $\lambda^{1}_0$ corresponds
to the one of the isolated dynamical unit. The remaining
variations $\xi_k$, k=1,2,...,N-1 are transverse to $\mathcal{M}$,
and describe the system's response to small deviations from the
synchronization manifold. Any deviation from the synchronization
manifold will be reflected in the growth of one or more of these
variations. The stability of the synchronized state is ensured if
arbitrary small transverse variations decay to zero. So, CS exists
if $\lambda^{1}_k<0$, for $k \geq 1$.


We also calculate the condition for the synchronization in the
network by using the Lyapunov spectra, calculated directly from
Eq. (\ref{mouk05}). Complete synchronization in the generalized
sense as defined in Refs. \cite{Boccaletti1,Boccaletti2} exists if
the second largest Lyapunov exponent is negative.

Due to the periodic potential in Eq. (\ref{mouk02}), the active
network in Eq. (\ref{mouk04}) is highly sensitive to initial
conditions.  As a consequence, networks whose elements have random
initial conditions that differ only slightly completely
synchronize for a coupling strength smaller than the coupling
strength needed to completely synchronize networks that have
elements whose initial conditions differ moderately. Often, the
network never complete synchronizes, and one can only have that
$|x_k-x_l|<\vartheta$, and so, the trajectory is never perfectly
along the synchronization manifold.  Even thought $\vartheta$
might be small, it is sufficiently large in order to mislead the
statement that complete synchronization appears by only checking
the conditional exponents.  This discrepancy is due to the fact
that, in this system, when the initial conditions are not too
close, the systems goes to different attractors and the
approximation made to obtain the conditional lyapunov exponents
[Eq. (\ref{mouk06})] is no longer completely valid, thought it
provides still approximate results. The effect of having nodes
with different initial conditions in the studied network is
similar to having networks with different parameters.

\begin{figure}[htbp]
  \epsfxsize=8cm \centerline{\epsfbox{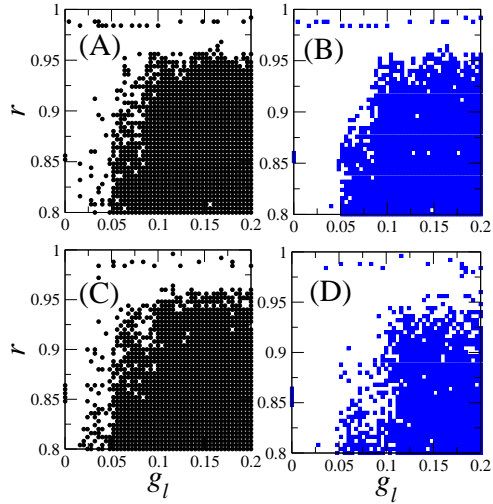}} \caption{
    Appearance of complete synchronization in a network of $N$=5
    diffusively coupled oscillators. Points represent $g_l$ and $r$
    values for which the conditional exponent $\lambda^{1}_1$ is
    negative (A,C) and for which the second largest Lyapunov exponent
    is negative (B,D). In (A,B), the initial conditions differ by at
    most 0.01 and in (C,D) the initial conditions differ by at most
    0.5.}
  \label{Fig.3}
\end{figure}

In Fig. \ref{Fig.3}, we show the parameter spaces (coupling $g_l$
vs. deformability parameter $r$) of the complete synchronization
regime. Points show the values of $g_l$ and $r$ for which all the
transversal ($k \ge 1$) conditional exponents are negative [Figs.
\ref{Fig.3}(A),(C)] or when the second largest Lyapunov exponent
becomes negative [Figs. \ref{Fig.3}(B),(D)].

When the initial conditions differ by no more than 0.01 [Figs.
\ref{Fig.3}(A-B)] the two conditions to predict complete
synchronization provide the same surface in the parameter space.
However, when these initial conditions differ by no more than 0.5
[Figs.  \ref{Fig.3}(C-D)], the conditional exponents predict the
appearance of complete synchronization for a coupling strength smaller
than the strength for which it really appears, as predicted by the
value of the second largest Lyapunov exponents [Figs.
\ref{Fig.3}(D)].


One can also observe from these figures that as the deformability
parameter increases, the system becomes more and more unstable.
When $r>9.5$, it is almost not possible to find complete
synchronization in the network for low values of coupling strength
$g_l$.  So, when the potential $V(P_i,r)$ has a flat bottom, the
particles are almost non-synchronizable in the network.
\section{Phase synchronization}\label{secaoIII}
Phase synchronization \cite{Kurths1,baptista:2006,pereira} is a
phenomenon defined by
\begin{equation}
|\phi_k - m \phi_l| \leq \epsilon, \label{phase_synchronization}
\end{equation}
\noindent where $\phi_k$ and $\phi_l$ are the phases of the nodes
$x_k$ and $x_l$ in the network [Eqs.(\ref{mouk03})] and
$m=\omega_l/\omega_k$, where $\omega_k$ and $\omega_l$ are the
average frequencies of oscillation of these nodes, and $\epsilon$
is a finite number. In this work, we have used in Eq.
(\ref{phase_synchronization}) $m=1$, which means that we search
for $\omega_k:\omega_l$=1:1 (rational) phase synchronization. If
another type of $\omega_k:\omega_l$-PS is present, the methods in
Refs. \cite{baptista:2006} can detect.

The phase $\phi$ is a function constructed on a good 2D subspace,
whose trajectory projection has proper rotation, i.e, it rotates
around a well defined center of rotation. Often, a good 2D
subspace is formed by the velocity space. In the oscillator
considered in this work, one can use the results of
\cite{pereira}, and define the phase of the oscillator $x_i$ in
Eqs. (\ref{mouk03}) as
\begin{equation}
\phi(t)=\int_0^t
\frac{\ddot{n}_i\dot{p}_i-\ddot{p}_i\dot{n}_i}{{(\dot{p}_i^2+\dot{n}_i^2)}}dt.
\label{phase_dxdy}
\end{equation}
However, the oscillators in Eqs. (\ref{mouk03}) for the considered
parameters have not a well defined phase, and even in a state
where complete synchronization is achieved, one cannot use Eq.
(\ref{phase_dxdy}) to verify whether PS exists.

In short, if PS exists, in a subspace, then the points obtained from
observations of the position of one node's trajectory at the time
another node makes any physical event do not visit the neighborhood of
a special curve $\Gamma$, in this subspace. A curve $\Gamma$ is
defined in the following way. Given a point $x_0$ in the attractor
projected onto the subspace of one oscillator where the phase is
defined, $\Gamma$ is the union of all points for which the phase,
calculated from this initial point $x_0$ reaches $n \langle r
\rangle$, with $n=1,2,3,\ldots,\infty$ and $\langle r \rangle$ a
constant, usually 2$\pi$. Clearly an infinite number of curves
$\Gamma$ can be defined.

\begin{figure}[htbp]
  \epsfxsize=7cm \centerline{\epsfbox{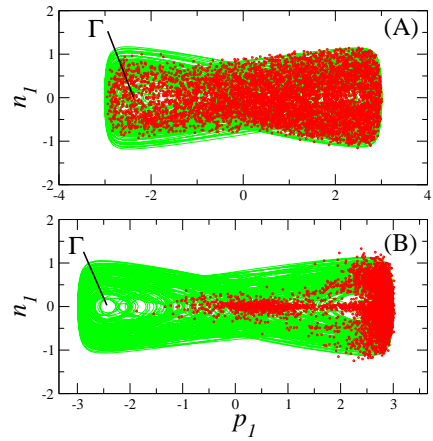}} \caption{The
    appearance of phase synchronization in two bidirectionally coupled
    oscillators. (A) There is no phase synchronization and the
    conditional observations are not localized with respect to the
    curve $\Gamma$ pictorially represented in the figure. (B) There is
    evidence of phase synchronization and the conditional observations
    are localized. Simulations are done considering initial conditions
    no more than 0.01 apart.}
\label{Fig.4}
\end{figure}

Formally, for non-coherent dynamical systems for which phase is
still not well defined, PS implies localization of the conditional
sets \cite{pereira}, but the contrary is not always true.
Therefore, finding localized sets should be considered a strong
evidence that PS exists.

As an example, consider Eqs. (\ref{mouk03}) with two coupled
oscillators, $r$=0.9, and $\phi=0.08$. For a small coupling
$g_l$=0.01, in Fig. \ref{Fig.4}(A), we show a situation that PS is
not present for $g_l$=0.01 and in Fig. \ref{Fig.4}(B), an evidence
that PS exists, for $g_l$=0.05. The curve $\Gamma$, a continuous
curve transversal to the trajectory, is pictorially represented by
the straight line $\Gamma$. In \ref{Fig.4}(A), the conditional observations
are not localized and thus there is no PS in this subspace. The
light gray line (green online) represents the attractor projection
on the subspace $(p_i,n_i)$ of the oscillator $x_1$, and filled
gray circles (red online) represent the points obtained from the
conditional observations of the oscillator $x_1$ whenever the
oscillator $x_2$ makes an event. An event is considered to be the
crossing of the trajectory to the line $n_2=0$, for $p_2>0$.

\begin{figure}[htbp]
  \epsfxsize=7cm \centerline{\epsfbox{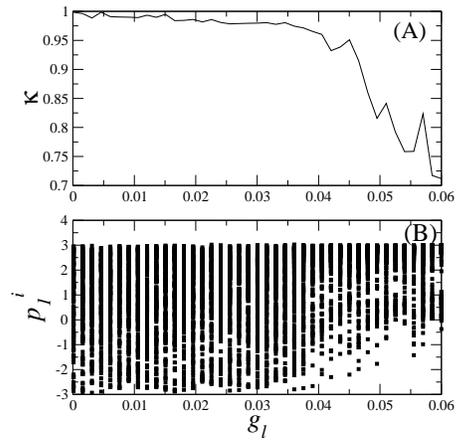}} \caption{The
    appearance of phase synchronization in two bidirectionally coupled
    oscillators. (A) Occupation of the conditional observations with
    respect to the attractor, $\kappa$, and in (B) the position
    variable $p_1^i$ when the oscillator $x_2$ makes the $i$th
    crossing with the section $n_2$=0, for $p_2>0$.}
\label{Fig.6}
\end{figure}

To have a general picture of when PS might appear in the two coupled
oscillators, we show in Fig. \ref{Fig.6}(A)
the quantity $\kappa$ with respect to $g_l$, defined as
\begin{equation}
\kappa=\frac{\max{(p_1^i) - \min{(p_1^i)}}}{\max{(p_1(t)) - \min{(p_1(t))}}}
\label{percentage}
\end{equation}
where $p_1^i$ represents the value of $p_1$ at the instant the
trajectory of oscillator $x_2$ makes an event. Therefore, $\kappa$
is related to how broad the conditional observations visit the
attractor. In Fig. \ref{Fig.6}(B) we show a few values of $p_1^i$
with respect to $g_l$. For $g_l \geq 0.06$, CS takes place.

For large networks composed of $N$ nodes, this analysis is straightforward and
PS between two nodes can be stated if the conditional observations realized in
one node, whenever the other node makes an event, produces a localized set.

\section{Information Transmission in the Network}\label{secaoIV}

\noindent In order to study the way information is transmitted in
active networks, we introduce quantities and terminologies that assist
us to better present our ideas and approaches.

The {\bf mutual information rate} (MIR) is the rate with which
information is being exchanged between two oscillation modes or
elements in the active network.

The {\bf channel capacity}, $\mathcal{C}_C$, is defined as the
maximal possible amount of information that two oscillation modes
or nodes within the network with a given topology can exchange, a
local measure that quantifies the point-to-point rate with which
information is being transmitted.


The {\bf Kolmogorov-Sinai entropy} offers an appropriate way of
obtaining the entropy production of a dynamical system. In chaotic
systems, the entropy equals the summation of all the positive
Lyapunov exponents (\cite{pesin}). Here, it provides a global
measure of how much information can be simultaneously transmitted
among all pairs of oscillation modes or nodes.  Therefore, the
KS-entropy, $H_{KS}$, of an active network, calculated for a given
coupling strength, bounds the MIR between two oscillation modes,
$I$, calculated for the same coupling strength. Thus,
\begin{equation}
I \leq  H_{KS} \label{limite}
\end{equation}

An active network is said to be {\bf self-excitable} ({\bf
  non-self-excitable}) when $\mathcal{C}_C$ $> H_{KS}^{(0)}$ (when
$\mathcal{C}_C$ $\leq H_{KS}^{(0)}$), with $H_{KS}^{(0)}$
representing the KS entropy of one of the $N$ elements forming the
active network, before they are coupled.

According to \cite{murilo}, the upper bound for the MIR between
two oscillation modes in an non-self-excitable active network,
denoted as $I$, can be calculated by
\begin{equation}
I^{k} \leq \lambda_{0}^1-\lambda_{k}^1 \label{mir}
\end{equation}
where $\lambda_{0}^1$ and $\lambda_{k}^1$ ($k=1,\ldots,N-1$) are
the positive largest conditional exponent \cite{pecora},
numerically obtained from Eq.  (\ref{mouk06}), with the
oscillators possessing equal initial conditions. $\lambda_{0}$
measures the exponential divergence of trajectories along the
synchronization manifold and $\lambda_{k}$ along the transversal
modes. The units used for the MIR is [bits/unit time], which can
be obtained by dividing Eq. (\ref{mir}) by $\log_e(2)$.

The networks as in Eq. (\ref{mouk04}) are predominantly of the
non-self-excitable type. Only for a very small coupling strength,
and a larger number of nodes, the network has a negligible
increase of the KS-entropy, which we will disregard.



\begin{figure}[htbp]
  \epsfxsize=8cm \centerline{\epsfbox{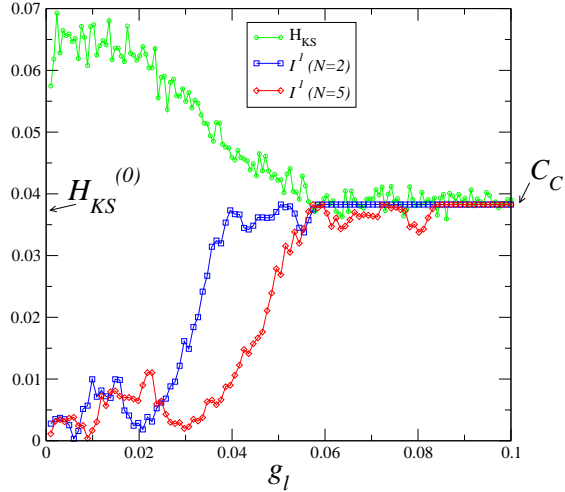}} \caption{[Color
    online] (green) Circles show the KS-entropy $H_{KS}$ and (blue)
    squares the MIR, $I^1$, for two bidirectional coupled oscillators.
    (red) Diamonds show $I^1$ for a network of $N$=5 diffusively
    coupled oscillators.}
\label{Fig.7}
\end{figure}

As can be seen from the $H_{KS}$ curve in Fig. \ref{Fig.7}, the
two coupled oscillators are of the non-self-excitable type, since
$H_{KS}^{(0)} = H_{KS}(g_l=0)/2$ which is approximately equal to
$\mathcal{C}_C$. In this figure, we also show the MIR exchanged between the
two coupled oscillators. As typically happens for non-excitable
networks, the channel capacity is reached when the network
complete synchronizes. Since the network is composed of two
bidirectionally coupled systems, the MIR between the only two
existing modes is actually the MIR between the two oscillators.

Comparing Figs. \ref{Fig.6}(A) and \ref{Fig.7}, one can see that there
is a direct relationship between synchronization and information. The
larger the amount of synchronization the larger the MIR, again another
typical character of non-excitable networks.

For larger networks with arbitrary topologies, the MIR between
oscillation modes is just a rescaled version of the MIR between
two coupled oscillators. Given that $g^{(2)}_l$ is the coupling
strength for which complete synchronization takes place in  two
coupled oscillators, and therefore this coupled system operates
with its channel capacity, the coupling strength for which
complete synchronization takes place in a whole network composed
of $N$ nodes with a certain topology is given by $g^{(N)}_l$
\begin{equation}
  \label{conditionMIR}
g^{(N)}_l=2\displaystyle\frac{g^{(2)}_l}{\gamma^1_1(N)}\:\:\:\:\:
\end{equation}
\noindent
 At the parameter
$g^{(N)}_l$, every pair of oscillators operate with the channel
capacity. Equation (\ref{conditionMIR}) means that having the
curve for the MIR for two coupled oscillators, the curve of the
MIR for larger networks is rescaled by the second largest
conditional Lyapunov exponent of the Laplacian matrix
$\gamma^1_1(N)$.

As an illustration of Eq. (\ref{conditionMIR}), we show in Fig.
\ref{Fig.7}, the MIR for a network composed of 5 oscillators
coupled diffusively. In this figure, we show the quantity $\langle
I \rangle$ defined as $\langle I \rangle=1/(N-1)\sum_1^{N-1} I^k$.
Note that even though $\langle I \rangle$ might change its values
according to the network topology and $N$, its maximal value is
bounded by the channel capacity, which do not depends on the $N$
and the topology, another typical characteristic of
non-self-excitable networks.

\section{Conclusion}\label{secaoV}

We study the relationship between synchronization and the rate with
which information is exchanged between nodes in a spatio-temporal
network which describes the dynamics of classical particles under a
substrate Remoissenet-Peyrard potential. In particular, we study
networks formed by Frenkel-Kontorova(FK) oscillators suffering the
action of harmonic interaction and non-sinusoidal substrate potential.

We show that such networks are predominantly of the
non-self-excitable type, i.e. as the coupling strength among the
nodes increases the KS-entropy decreases. Other additional
characteristics of non-self-excitable networks are: the mutual
information rate (MIR) and the synchronization level increase
simultaneously as the KS-entropy decreases; the channel capacity,
the maximal of the MIR, is achieved for the same coupling strength
for which complete synchronization appears.

We have overcome two difficulties concerning the detection of
phase and complete synchronization in this complex spatio-temporal
network. Even though the phase dynamics of each oscillator is not
well defined, we have implement a technique which allows to
evidence the presence of phase synchronization, by detecting the
presence of localized sets obtained by the conditional
observations. The more localized the sets are (which implies
larger amount of phase synchrony) the larger the MIR. Concerning
complete synchronization, we show that the master stability
equation which provides the stability of the normal transversal
modes (providing conditions to state complete synchronization)
should be used with caution in such a network. The reason is that
the final state is highly dependent on the initial conditions, a
consequence of the spatio character provided by the potential. For
that reason, in case the nodes have sufficiently different initial
conditions, one should only state complete synchronization using
the master stability equation in an approximate sense. A more
rigorous condition to state complete synchronization is provided
by the verification that the second largest Lyapunov exponent is
negative.

Finally, we have shown how one can calculate the MIR between
oscillation modes in larger networks with different topologies
using as the only input information the curve of the MIR with
respect to the coupling strength for two bidirectionally coupled
oscillators. Having the curve for the MIR for two coupled
oscillators, the curve of the MIR for larger networks is rescaled
by the second largest conditional Lyapunov exponent of the
Laplacian matrix of the larger network, the matrix that describes
the way the nodes are connected in the network. That enables one
to construct larger networks based on the dynamical
characteristics of only two coupled oscillators.
\section*{Acknowledgments}
Both authors acknowledge the wonderful time spend in the Max Planck Institute
for the Physics of Complex Systems (MPIPKS) and thank the financial support
provided by this Institute. We also thank Sara P. Garcia for a critical
reading of the manuscript.
%

%
\end{document}